\documentclass[twocolumn,secnumarabic,amssymb,nobibnotes,aps,prd,showpacs]{{revtex4-1}}
\usepackage{amsmath}
\usepackage{amssymb}
\usepackage[dvipdfm]{epsfig}
\usepackage{graphicx}
\begin{document}
\title {
Unified Model of Temperature Dependence of Core Losses in Soft 
Magnetic Materials Exposed to Nonsinusoidal
Flux Waveforms and DC Bias Condition}
\author{Adam Ruszczyk}
\email[e-mail: ]{adam.ruszczyk@pl.abb.com}
\affiliation{ ABB Corporate Research, Starowi\'{s}lna 13a, 31-038 Krak\'{o}w, Poland,}

\author{Krzysztof Sokalski}
\email[e-mail: ]{sokalski\_krzysztof@o2.pl}
\affiliation{Institute of Computer Science, 
Cz\c{e}stochowa University of Technology,
Al. Armii Krajowej 17, 42-200 Cz\c{e}stochowa, Poland}



\begin{abstract}
  Assuming that Soft Magnetic Material is a Complex System and expressing this feature by scaling invariance of the power loss characteristic, the unified model of the temperature dependence of Core Losses in Soft Magnetic Materials Exposed to Nonsinusoidal Flux Waveforms and DC Bias Condition has been constructed. 
 In order to verify this achievement the appropriate measurement data concerning power losses and the all independent variables have been collected. The model parameters have been estimated and  the power losses modeling has been performed. Comparison of the experimental values of power losses with their calculated values has showed  good agreement.
\end{abstract}
\pacs{75.50.-y, 61.85.+p}
\maketitle
{\small The following article has been submitted to Applied Physics Letters.  If it is published, it will be found online at http://apl.aip.org}
\section*{Introduction}

 Application  of soft magnetic materials in electronic devices requires knowledge about losses under different conditions of exposition: sinusoidal and nonsinusoidal flux waveforms of different shapes, with and without DC bias condition. During the two last decades the two classes of core loss' models have been elaborated.  The first class consists of models which are based on the Steinmetz Equation 
\citep{bib:Steinmetz}, \citep{bib:Albach}, \citep{bib:Reinert}, \citep{bib:Li},\citep{bib:Venka}, \citep{bib:Boss1}, \citep{bib:Ecklebe}, \citep{bib:Ecklebe1},\citep{bib:FFio}. However the second class is based on the assumption that the shape of the waveform does not matter and as a result only look at peaks \citep{bib:Sokal1},\citep{bib:Sokal2},\citep{bib:Sokal3},\citep{bib:Ruszcz}  and \citep{bib:Cale}.
Non of them presents satisfactory algoritm enabling us to calculate of core losses v.s. temperature of  sample with and without presence of conditions for exposition mentioned above. Therefore, this paper is devoted to solution of this problem. 
\section{Scaling and Unified Core Loss Model}
On the base of our recent papers \citep{bib:Sokal1},\citep{bib:Ruszcz} we derive the unified model of the total core loss versus the four independent variables: $f$-frequency, $\triangle B$-pik to pik magnetic induction, $H_{DC}$-DC bias and $T$-temperature:
\begin{equation}
\label{T1}
P_{tot}=F(f,\triangle B,H_{DC},T).
\end{equation}
In order to apply scaling to (\ref{T1}) the right hand side has to be homogeneous function in general sense. This assumption has to be satisfied both, by the experimental data and by the mathematical model. However,
according to results of researches  presented in \citep{bib:ABB1}, (\ref{T1}) and measurement data formed by the action of DC-bias are not uniform in the required sense. This problem we have solved in the previous paper \citep{bib:Ruszcz} by using the method invented by Van den Bossche et al. \citep{bib:Boss1}. They have  mapped the DC-bias into primary magnetization curve. Using their idea we have used the following mapping:
\begin{equation}
\label{T2}
H_{DC}\rightarrow [M_{0},M_{1},M_{2},M_{3}],
\end{equation}
where $M_{i}=tanh{(H_{DC}\cdot c_{i})}$ and $c_{i}$ are free parameters to be determined from the experimental data. The number of $M_{i}$ components is optional.
The introduced mapping (\ref{T2}) enables us to write down the following condition for $P_{tot}(f,\triangle B, [M_{0},M_{1},M_{2},M_{3}],T)$ to be a homogenous function in general sense:
\begin{eqnarray}
\exists \{a,b,c,d,g\}\in {\bf R^{5}}: \forall \lambda\in \bf{R_{+}}\nonumber\\
P_{tot}(\lambda^{a}f,\lambda^{b}(\triangle B), \lambda^{c}[M_{0},M_{1},M_{2},M_{3}],\lambda^{d}T)=\nonumber\\
\lambda^{g}P_{tot}(f,\triangle B, [M_{0},M_{1},M_{2},M_{3}],T).\label{T3}
\end{eqnarray}
Substituting for  $\lambda$ the following expression: \\$\lambda=(\triangle B)^{-1/b}$ we derive the most general form for $P_{tot}$ which satisfies (\ref{T3}):
\begin{equation}
\label{T4}
P_{tot}=(\triangle B)^{\beta}\,F\left(\frac{f}{(\triangle B)^{\alpha}}, \frac{[M_{0},M_{1},M_{2},M_{3}]}{(\triangle B)^{\gamma}} ,\frac{T}{(\triangle B)^{\delta}} \right),
\end{equation}
 where, $\alpha=\frac{a}{b},\beta=\frac{g}{b},\gamma=\frac{c}{b},\delta=\frac{d}{b}$ and $F(\cdot,\cdot,\cdot)$ is an arbitrary function to be determined. 
\section{The modeling of $F(\cdot,\cdot,\cdot)$ }
In order to determine $F(\cdot,\cdot,\cdot)$  we assume its form to be factorable:
\begin{eqnarray}
F\left(\frac{f}{(\triangle B)^{\alpha}}, \frac{[M_{0},M_{1},M_{2},M_{3}]}{(\triangle B)^{\gamma}} ,\frac{T}{(\triangle B)^{\delta}} \right)=\nonumber\\
\Phi\left(\frac{f}{(\triangle B)^{\alpha}},\frac{[M_{0},M_{1},M_{2},M_{3}]}{(\triangle B)^{\gamma}}\right)\,\Theta \left(\frac{T}{(\triangle B)^{\delta}}\right).\label{scal2}
\end{eqnarray}
$\Phi(\cdot,\cdot)$ is a version of very well working model function derived in \citep{bib:Ruszcz}:
\begin{eqnarray}
\Phi(\frac{f}{(\triangle B)^{\alpha}},H_{DC})=\Sigma_{i=1}^{4}\Gamma_{i}\left (\frac{f}{(\triangle B)^{\alpha}}\right)^{i\,(1-x)}+\nonumber\\\label{applic2}
\Sigma_{i=0}^{3}\Gamma_{i+5}\left (\frac{f}{(\triangle B)^{\alpha}}\right)^{(i+y)(1-x)}\frac{tanh(H_{DC}\cdot c_{i})}{(\triangle B)^{\delta}}.
\end{eqnarray}
Basing on  some computer experiments we have selected for $\Theta(\cdot)$ the following Pad\'{e} approximant \citep{bib:pade}:
\begin{equation}
\label{scal4}
\Theta=\left (\frac{\psi_{0}+\theta\,(\psi_{1}+\theta\,\psi_{2})}{1+\theta\,(\psi_{3}+\theta\,\psi_{4})}\right)^{1-z},
\end{equation}
where $\theta=\frac{T+\tau}{\Delta B^{\gamma}}$, $T^{\circ}C$ is measured temperature, $\tau$ and $z$ are tuning parameters, $\psi_{i}$ are Pad\'{e} expansion coefficients.
\section{Experimental Data, Estimations of Parametr's and Modeling}

\begin{table*}
\caption{Selected 60 records of the measurement data of SIFERRIT N87 }
\label{APPENDIX}
\begin{center}
\begin{tabular}{cccccccccc}\hline
 $T[^oC]$&$\triangle B[T]$	&	$f[kHz]$&	$H_{DC}[\frac{A}{m}]$	&	$P_{tot}	[\frac{W}{m^{3}}]$&$T[^oC]$&$\triangle B[T]$	&	$f[kHz]$&	$H_{DC}[\frac{A}{m}]$	& $P_{tot}[\frac{W}{m^{3}}]$\\\hline
	28,1&	0,395&	1&	8,634&	   4064,3&	28,1&	0,391&	1&	20,146&	4469,0\\
	28,1&	0,374&	1&	60,634&	6332,4&28,3&	0,351&	1&	86,651&	6463,6\\
	17,7&	0,398&	2&	7,8014&	9452,1&	17,8&	0,398&	2&	20,555&	10663,8\\
	18,9&	0,396&	2&	35,583&	12745,8& 18,5&	0,377&	2&	89,240&	16015,6	\\
	26,2&	0,400&	5&	6,570&	    21131,3&	26,4&	0,400&	5&	17,820&	23110,0\\
	26,5&	0,398&	5&	33,230&	28057,3&27,1&	0,386&	5&	89,400&	35209,8\\
		28,4&	0,401&	10&	5,892&	    41549,0&	28,6&	0,401&	10&	17,477&	45257,9\\
	28,8&	0,400&	10&	31,820&	54650,9&29,7&	0,393&	10&	73,960&	63821,6\\
	30,8&	0,386&	10&	105,00&	64632,1&	28,4&	0,490&	1&	11,694&	6611,0\\
	28,4&	0,488&	1&	24,299&	7196,0&28,4&	0,451&	1&	78,390&	8771,6\\
	19,1&	0,497&	2&	10,120&	15234,1&	19,2&	0,496&	2&	23,718&	16781,0\\
    19,3&	0,485&	2&	54,63&	19235,9&19,8&	0,475&	2&	76,86&	20100,2\\	
	27,7&	0,502&	5&	8,92&	34634,8&	27,4&	0,503&	5&	15,02&	36195,2\\
	27,7&	0,501&	5&	21,5&	37496,6&	28,6&	0,496&	5&	47,5&	41259,7\\	
	31,7&	0,499&	10&	20,52&	71226,8&	32,15&	0,494&	10&	45,04&	76876,5\\
	32,6&	0,487&	10&	67,14&	80858,2&	28,5&	0,588&	1&	14,42&	10042,9\\
	28,7&	0,561&	1&	57,97&	11239,6&	28,7&	0,541&	1&	78,08&	11255,7\\
	29,1&	0,58&	2&	12,82&	19689,9&	28,7&	0,576&	2&	54,36&	22043,0\\
	30,1&	0,592&	5&	42,4&	52126,7&	31,1&	0,599&	10&	10,29&	92648,6\\
	31,3&	0,595&	10&	31,23&	96446,4& 	28,9&	0,684&	1&	22,05&	14150,5\\
	28,1&	0,389&	1&	33,507&	5358,8& 28,4&	0,346&	1&	91,066&	6376,4\\
    18,2&	0,386&	2&	68,034&	15049,1&18,7&	0,367&	2&	110,59&	16027,7\\
	29&	0,669&	1&	41,33&	14417,5&34,7&	0,586&	10&	61,25&	96583,3\\
    30,2&	0,616&	5&	36,05&	54344,9&28,7&	0,586&	2&	33,49&	21002,2\\
  	28,5&	0,580&	1&	36,01&	10790,0& 42,1&	0,496&	50&	47,53&	289491,2\\
31,5&	0,499&	10&	7,57&	65879,7&28,1&	0,500&	5&	31,42&	39530,2\\
20,2&	0,469&	2&	87,44&	20547,5&19,7&	0,480&	2&	68,36&	20073,3\\
	28,5&	0,443&	1&	85,100&	8702,4&28,3&	0,473&	1&	54,300&	8296,5\\
	30,2&	0,387&	10&	99,190&	64410,1&	29,2&	0,396&	10&	61,172&	62814,4\\
	27,5&	0,386&	5&	97,779&	35945,6&	26,8&	0,394&	5&	58,800&	32614,3\\
\end{tabular}
\end{center}
\end{table*} 

\begin{figure}
\includegraphics [angle=0, width=8 cm]{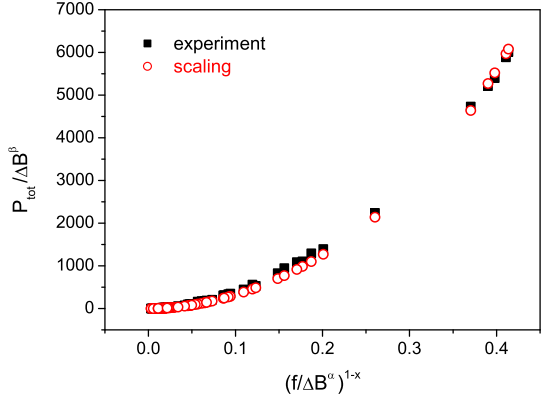}
\caption{ Projection of the measurement points and  the scaling theory points (\ref{scal2})-(\ref{scal4})  in $[f/( \triangle B^{\alpha})^{(1-x)},P_{tot}/(\triangle B^{\beta})]$ plane.}
\label{Fig.1a}
\end{figure}
\begin{figure}
\includegraphics [angle=0, width=8 cm]{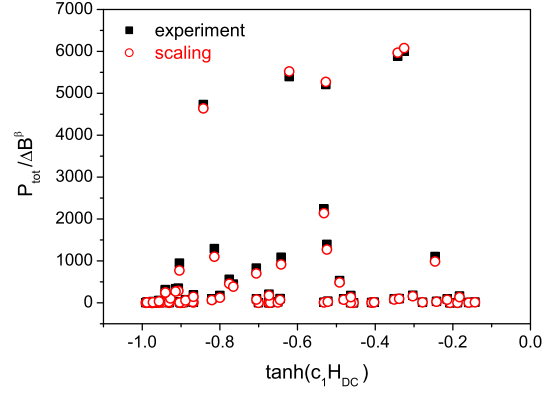}
\caption{ Projection of the measurement points and  the scaling theory points (\ref{scal2})-(\ref{scal4})  in $[tanh(H\,c_{1},P_{tot}/(\triangle B)^{\beta}]$ plane.}
\label{Fig.2a}
\end{figure}
\begin{figure}
\includegraphics [angle=0, width=8 cm]{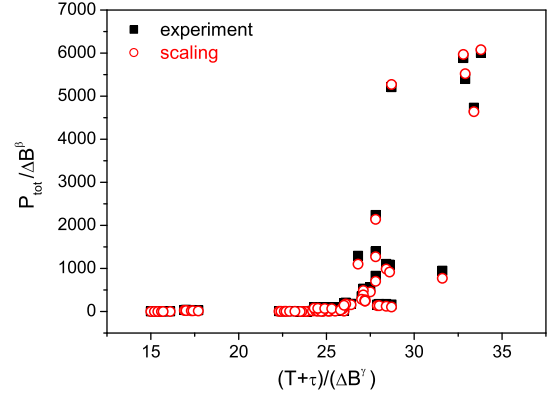}
\caption{ Projection of the measurement points and  the scaling theory points (\ref{scal2})-(\ref{scal4}) in $[\frac{T+\tau}{(\triangle B)^{\gamma}},P_{tot}/(\triangle B)^{\beta}]$ plane.}
\label{Fig.3a}
\end{figure}

The B-H Loop measurements have been performed for SIFERRIT N87. The Core Losses per unit volume have been calculated as the
enclosed area of the B-H loop, multiplied by the frequency f.
The following factors influence the accuracy of measurements:
1) Phase Shift Error of Voltage and Current $<4\%$, 2)
Equipment Accuracy $<5,6\%$, 3) Capacitive Couplings
negligible (capacitive currents are relatively lower compared to
inductive currents), and 4) Temperature $<4\%$. For details of
the applied measurement method and the errors of the relevant
factors we refer to \citep{bib:Ecklebe}, \citep{bib:Ecklebe1}. 
The parameter values of (\ref{T4})-(\ref{scal4}) have been estimated by minimization of $\chi^{2}$ using the Simplex
method of Nelder and Mead \citep{bib:pade} and the our experimental data. The measurement series consists of $60$ points, see TABLE \ref{APPENDIX}. 
Standard deviation per point is equal to $15[\frac{W}{m^{3}T^{\beta}}]$
 Applying the formulae (\ref{T4})-(\ref{scal4}) and the
estimated parameter values TABLE \ref{Tab1} we have drawn the three scatter plots Fig. \ref{Fig.1a}, Fig. \ref{Fig.2a} and Fig. \ref{Fig.3a},  which
compare estimated points with the experimental ones in the three projections, respectively. Note that, in order to prevent
generation of large numbers in the estimation process the unit
of frequency was kHz while other magnitudes were expressed
in SI unit system.

\begin{table*}
\begin{center}
\caption{The set of  estimated model's parameters of (\ref{T4})-(\ref{scal4}) for $\delta=0 $}
\label{Tab1}
\begin{tabular}{cccccccc}\hline
$\alpha$&$\beta$&$x$&$\Gamma_{1}$&$\Gamma_{2}$&$\Gamma_{3}$&$\Gamma_{4}$&$\Gamma_{5}$\\\hline
${\small -11,628}$&	-8,6382&	0,52629	&-1,4083	&739,55	&1253,4	&4238,5	&0,12264\\\hline\hline
$\Gamma_{6}$ & $\Gamma_{7}$&$\Gamma_{8}$& y & $\psi_{3}$ & $c_{3}$ & $\psi_{4}$  & $\psi_{5}$\\\hline
-30,972	&-51,869&-4201,45&	0,28877&14,4558&	0,1648	&-1,27E-01&	0,28302\\\hline\hline
 $c_{2}$&$\tau$& $\gamma$ & $\psi_{2}$&$\psi_{1}$&$c_{1}$& $c_{0}$ & z\\\hline
-0,1808&	7,77E-02&	-0,17954&	2,3966	&-0,8993&-2,44E-02	&-0,4877&	4,84E-02\\\hline
\end{tabular}
\end{center}
\end{table*}
\section{Conclusions}
Efficiency of the scaling in solving problems concerning of power losses in Soft Magnetic Materials has been confirmed all ready in the recent papers \citep{bib:Sokal1}-\citep{bib:Ruszcz}. 
However, this paper is the first one which presents application of scaling in modeling of temperature dependence of the core loss. The presented method is universal, which means that it works for wide spectrum of expositions and different soft magnetic materials. Moreover the presented model formulae (\ref{T4})-(\ref{scal4}) are not closed and can be adapted for a current problem by fitting 
the forms of both factors $\Phi$ and $\Theta$. At the end one must say that success in applying the scaling depends on property of data. The data must obey the scaling. 

\section*{}

\bibliographystyle{plainnat}
\end{document}